# A Computer Verified Theory of Compact Sets*

by Russell O'Connor

Institute for Computing and Information Science
Faculty of Science
Radboud University Nijmegen

*Email:* r.oconnor@cs.ru.nl

**Abstract**

Compact sets in constructive mathematics capture our intuition of what computable subsets of the plane (or any other complete metric space) ought to be. A good representation of compact sets provides an efficient means of creating and displaying images with a computer. In this paper, I build upon existing work about complete metric spaces to define compact sets as the completion of the space of finite sets under the Hausdorff metric. This definition allowed me to quickly develop a computer verified theory of compact sets. I applied this theory to compute provably correct plots of uniformly continuous functions.

## 0  Licence

## 1  Introduction

How should we define what computable subsets of the plane are? Sir Roger Penrose ponders this question at one point in his book "The Emperor's New Mind" [9]. Requiring that subsets be decidable is too strict; determining if a point lies on the boundary of a set is undecidable in general. Penrose gives the unit disc, $\{(x,y)|x^2+y^2 \leq 1\}$, and the epigraph of the exponential function, $\{(x,y)|\exp(x) \leq y\}$, as examples of sets that intuitively ought to be considered computable [2]. Restricting one's attention to pairs of rational or algebraic numbers may work well for the unit disc, but the boundary of the epigraph of the exponential function contains only one algebraic point. A better definition is needed.

To characterize computable sets, we draw an analogy with real numbers. The computable real numbers are real numbers that can be effectively approximated to arbitrary precision. The approximations are usually rational numbers or dyadic rational numbers. We can define computable sets in a similar way.

We need a dense subset of sets that have finitary representations. In the case of the plane, the simplest candidate is the finite subsets of $\mathbb{Q}^2$. Again, $\mathbb{Q}$ could be replaced with the dyadic rationals. How do we measure the accuracy of an approximation? Distances between subsets can be defined by the Hausdorff metric (section 3).

To construct the real numbers, we complete the rational numbers. By reasoning constructively (section 2), the real numbers generated are always computable. Completing the finite subsets of $\mathbb{Q}^2$ with the Hausdorff metric yields the compact sets (section 5). By reasoning constructively, the generated compact sets are always computable!

---

The unit disc is constructively compact; it can be effectively approximated with finite sets. When a computer attempts to display the unit disc, only a finite set of the pixels can be shown. So instead of displaying an ideal disc, the computer displays a finite set that approximates the disc. This is the key criterion that Penrose's examples enjoy. They can be approximated to arbitrary precision and displayed on a raster.

Technically the epigraph of the exponential function is not compact; however, it is locally compact. One may wish to consider constructive locally compact sets to be computable. This would mean that any finite region of a computable set has effective approximations of arbitrary precision.

This definition of constructively compact sets has been formalized in the Coq proof assistant [11]. Approximations of compact sets can be rasterized and displayed inside Coq (section 6). For example, figure 1 shows a theorem in Coq certifying that a plot is close the exponential function. The plot itself is computed from the definition of the graph of the exponential function.

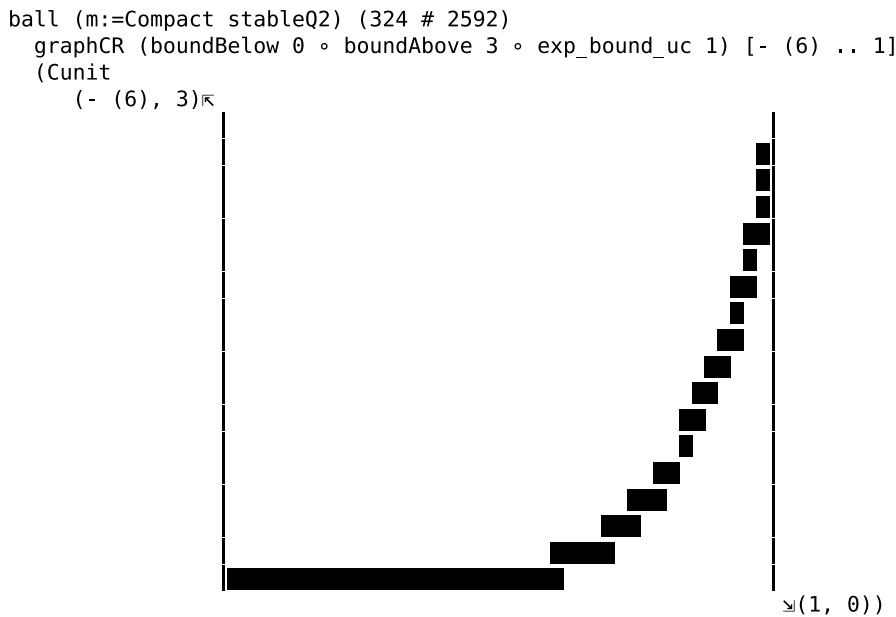

**Figure 1.** A theorem in Coq stating that a plot on a 42 by 18 raster is close to the graph of the exponential function on $[-6, 1]$.

The standard definition of computable sets used in computable analysis says that a set is computable if the distance to the set is a computable real-valued function. This definition is equivalent to our definition using computable approximations (although, this has not been verified in Coq). However, I believe defining computable sets by effective approximations of finite sets more accurately matches our intuition about sets that can be drawn by a computer.

## 2 Constructive Mathematics

Usually constructive logic is presented as a restriction of classical logic where proof by contradiction and the law of the excluded middle are not allowed. While this is a valid point of view, constructive logic can instead be presented as an extension of classical logic.

Consider formulas constructed from universal quantification ($\forall$), implication ($\Rightarrow$), conjunction ($\wedge$), true ($\top$), false ($\bot$), and equality for natural numbers ($=_\mathbb{N}$). Define negation by $\neg \varphi := \varphi \Rightarrow \bot$. One can (constructively) prove $\neg\neg\varphi \Rightarrow \varphi$ holds for any formula $\varphi$ generated from this set of connectives by induction on the structure of $\varphi$ because the atomic formulas—which in this case are equalities on $\mathbb{N}$—are decidable. Thus, one can deduce classical results with constructive proofs for formulas generated from this restricted set of connectives.

This set of connectives is not really restrictive because it can be used to define the other connectives. One can define the classical disjunction ($\tilde{\vee}$) by $\varphi \tilde{\vee} \psi := \neg(\neg\varphi \wedge \neg\psi)$. Similarly, one can define the classical existential quantifier ($\tilde{\exists}$) by $\tilde{\exists} x. \varphi(x) := \neg\forall x. \neg\varphi(x)$. With this full set of connectives, one can produce classical mathematics. The law of the excluded middle ($\varphi \tilde{\vee} \neg\varphi$) has a constructive proof when the classical disjunction is used.

Given this presentation of classical logic, we can extend the logic by adding two new connectives, the constructive disjunction ( $\vee$ ) and the constructive existential ($\exists$). These new connectives come equipped with their constructive rules of inference (given by natural deduction) [12]. These constructive connectives are slightly stronger than their classical counterparts. Constructive excluded middle ($\varphi \vee \neg\varphi$) cannot be deduced in general, and our inductive argument that $\neg\neg\varphi \Rightarrow \varphi$ holds no longer goes through if $\varphi$ uses these constructive connectives.

We wish to use constructive reasoning because constructive proofs have a computational interpretation. A constructive proof of $\varphi \vee \psi$ tells which of the two disjuncts hold. A proof of $\exists n \colon \mathbb{N}. \varphi(n)$ gives an explicit value for $n$ that makes $\varphi(n)$ hold. Most importantly, we have a functional interpretation of $\Rightarrow$ and $\forall$. A proof of $\forall n \colon \mathbb{N}. \exists m \colon \mathbb{N}. \varphi(n, m)$ is interpreted as a function with an argument $n$ that returns an $m$ paired with a proof of $\varphi(n, m)$.

The classical fragment also admits this functional interpretation, but formulas in the classical fragment typically end in ... $\Rightarrow \bot$. These functions take their arguments and return a proof of false. Of course, there is no proof of false, so it must be the case that the arguments cannot simultaneously be satisfied. Therefore, these functions can never be executed. In this sense, only trivial functions are created by proofs of classical formulas. This is why constructive mathematics aims to strengthen classical results. We wish to create proofs with non-trivial functional interpretations.

From now on, I will leave out the word "constructive" from phrases like "constructive disjunction" and "constructive existential" and simply write "disjunction" and "existential". This follows the standard practice in constructive mathematics of using names from classical mathematics to refer to some stronger constructive notion. I will explicitly use the word "classical" when I wish to refer to classical concepts.

### 2.1 Dependently Typed Functional Programming

This functional interpretation of constructive deductions is given by the Curry-Howard isomorphism [12]. This isomorphism associates formulas with dependent types, and proofs of formulas with functional programs of the associated dependent types. For example, the identity function $\lambda x \colon A. x$ of type $A \Rightarrow A$ represents a proof of the tautology $A \Rightarrow A$. Table 1 lists the association between logical connectives and type constructors.

| Logical Connective | Type Constructor |
|---|---|
| implication: $\Rightarrow$ | function type: $\Rightarrow$ |
| conjunction: $\wedge$ | product type: $\times$ |
| disjunction: $\vee$ | disjoint union type: $+$ |
| true: $\top$ | unit type: () |
| false: $\bot$ | void type: $\emptyset$ |
| for all: $\forall x. \varphi(x)$ | dependent function type: $\Pi x. \varphi(x)$ |
| exists: $\exists x. \varphi(x)$ | dependent pair type: $\Sigma x. \varphi(x)$ |

**Table 1.** The association between formulas and types given by the Curry-Howard isomorphism

In dependent type theory, functions from values to types are allowed. Using types parametrized by values, one can create dependent pair types, $\Sigma x \colon A. \varphi(x)$, and dependent function types, $\Pi x \colon A. \varphi(x)$. A dependent pair consists of a value $x$ of type $A$ and an value of type $\varphi(x)$. The type of the second value depends on the first value, $x$. A dependent function is a function from the type $A$ to the type $\varphi(x)$. The type of the result depends on the value of the input.

The association between logical connectives and types can be carried over to constructive mathematics. We associate mathematical structures, such as the natural numbers, with inductive types in functional programming languages. We associate atomic formulas with functions returning types. For example, we can define equality on the natural numbers, $x =_{\mathbb{N}} y$, as a recursive function:

$$\begin{aligned} 0 =_{\mathbb{N}} 0 &:= \top \\ Sx =_{\mathbb{N}} 0 &:= \bot \\ 0 =_{\mathbb{N}} Sy &:= \bot \\ Sx =_{\mathbb{N}} Sy &:= x =_{\mathbb{N}} y \end{aligned}$$

One catch is that general recursion is not allowed when creating functions. The problem is that general recursion allows one to create a fixpoint operator $\text{fix}: (\varphi \Rightarrow \varphi) \Rightarrow \varphi$ that corresponds to a proof of a logical inconsistency. To prevent this, we allow only well-founded recursion over an argument with an inductive type. Because well-founded recursion ensures that functions always terminate, the language is not Turing complete. However, one can still express fast growing functions like the Ackermann function without difficulty [12].

Because proofs and programs are written in the same language, we can freely mix the two. For example, in my previous work [7], I represent the real numbers by the type

$$\exists f: \mathbb{Q}^+ \Rightarrow \mathbb{Q}. \forall \varepsilon_1\, \varepsilon_2. |f(\varepsilon_1) - f(\varepsilon_2)| \leq \varepsilon_1 + \varepsilon_2. \tag{1}$$

Values of this type are pairs of a function $f: \mathbb{Q}^+ \Rightarrow \mathbb{Q}$ and a proof of $\forall \varepsilon_1\, \varepsilon_2. |f(\varepsilon_1) - f(\varepsilon_2)| \leq \varepsilon_1 + \varepsilon_2$. The idea is that a real number is represented by a function $f$ that maps any requested precision $\varepsilon: \mathbb{Q}^+$ to a rational approximation of the real number. Not every function of type $\mathbb{Q}^+ \Rightarrow \mathbb{Q}$ represents a real number. Only those functions that have coherent approximations should be allowed. The proof object paired with $f$ witnesses the fact that $f$ has coherent approximations. This is one example of how mixing functions and formulas allows one to create precise datatypes.

## 2.2 Notation

I will use the functional style of defining multivariate functions with Curried types. A binary function will have type $X \Rightarrow Y \Rightarrow Z$ instead of $X \wedge Y \Rightarrow Z$ ($\Rightarrow$ is taken to be right associative). To ease readability, I will still write binary function application as $f(x, y)$, even though it should really be $f(x)(y)$.

Anonymous functions are written using lambda expressions. A function on natural numbers that doubles its input is written $\lambda x: \mathbb{N}. 2\, x$. The type of the parameter will be omitted when it is clear from context what it should be.

The type of propositions is $\star$. Predicates are represented by functions to $\star$. These predicates are often used where power sets are used in classical mathematics. The type $X \Rightarrow \star$ can be seen as the power set of $X$. I will often write $x \in \mathcal{A}$ in place of $\mathcal{A}(x)$ when $\mathcal{A}: X \Rightarrow \star$ and $x: X$.

The notation $x \in l$ is also used when $l$ is a finite enumeration (section 4). Also $x \in \mathcal{S}$ will be used when $\mathcal{S}$ is a compact set (section 5). The types will make it clear what the interpretation of $\in$ should be.

I will use shorthand to combine membership with quantifiers. I will write $\forall x \in \mathcal{A}. \varphi(x)$ for $\forall x. x \in \mathcal{A} \Rightarrow \varphi(x)$, and $\exists x \in \mathcal{A}. \varphi(x)$ will mean $\exists x. x \in \mathcal{A} \wedge \varphi(x)$.

Quotient types are not used in this theory. In place of quotients, setoids are used. A *setoid* is a dependent record containing a type $X$ (its carrier), a relation $\asymp: X \Rightarrow X \Rightarrow \star$, and a proof that $\asymp$ is an equivalence relation. When we define a function on setoid, we usually prove it is *respectful*, meaning it respects the setoid equivalence relations on its domain and codomain. Respectful functions will also be called *morphisms*.

I will often write $f(x)$ when $f$ is a record (or existential) with a function as its carrier (or witness) and leave implicit the projection of $f$ into a function.

# 3 Metric Spaces

Traditionally, a metric space is defined as a set $X$ with a metric function $d\colon X \times X \Rightarrow \mathbb{R}^{0+}$ satisfying certain axioms. The usual constructive formulation requires $d$ be a computable function. In my previous work [7], I have found it useful to take a more relaxed definition for a metric space that does not require the metric be a function. Instead, I represent the metric via a respectful ball relation $B\colon \mathbb{Q}^+ \Rightarrow X \Rightarrow X \Rightarrow \star$ satisfying five axioms:

1. $\forall x\,\varepsilon.\, B_\varepsilon(x,x)$
2. $\forall x\,y\,\varepsilon.\, B_\varepsilon(x,y) \Rightarrow B_\varepsilon(y,x)$
3. $\forall x\,y\,z\,\varepsilon_1\,\varepsilon_2.\, B_{\varepsilon_1}(x,y) \Rightarrow B_{\varepsilon_2}(y,z) \Rightarrow B_{\varepsilon_1+\varepsilon_2}(x,z)$
4. $\forall x\,y\,\varepsilon.\, (\forall \delta.\, \varepsilon < \delta \Rightarrow B_\delta(x,y)) \Rightarrow B_\varepsilon(x,y)$
5. $\forall x\,y.\, (\forall \varepsilon.\, B_\varepsilon(x,y)) \Rightarrow x \asymp y$

The ball relation $B_\varepsilon(x,y)$ expresses that the points $x$ and $y$ are within $\varepsilon$ of each other. I call this a ball relationship because the partially applied relation $B_\varepsilon(x)\colon X \Rightarrow \star$ is a predicate that represents the ball of radius $\varepsilon$ around the point $x$. The first two axioms are reflexivity and symmetry of the ball relationship. The third axiom is a version of the triangle inequality.

The fourth axiom states that the balls are closed balls. Closed balls are used because being closed is usually a classical formula. This means they can be ignored during computation because they have no computational content [4]. We want to minimize the amount of computation needed to get our constructive results.

The fifth axiom states the identity of indiscernibles. This means that if two points are arbitrarily close together then they are equivalent. The reverse implication follows from the reflexivity axiom and the fact that $B$ is respectful. In some instances, axiom 5 can be considered as the definition of $\asymp$ on $X$.

For example, $\mathbb{Q}$ can be equipped with the usual metric by defining the ball relation as

$$B_\varepsilon^{\mathbb{Q}}(x,y) := |x-y| \leq \varepsilon.$$

This definition satisfies all the required axioms.

## 3.1 Uniform Continuity

We are interested in the category of metric spaces with uniformly continuous functions between them. A function $f\colon X \Rightarrow Y$ between two metric spaces is *uniformly continuous with modulus* $\mu_f\colon \mathbb{Q}^+ \Rightarrow \mathbb{Q}^+$ if

$$\forall x_1\,x_2\,\varepsilon.\, B^X_{\mu_f(\varepsilon)}(x_1,x_2) \Rightarrow B^Y_\varepsilon(f(x_1),f(x_2)).$$

We call a function *uniformly continuous* if it is uniformly continuous with some modulus. We use notation $X \to Y$ with a single bar arrow to denote the type of uniformly continuous functions from $X$ to $Y$. This record type consists of three parts, a function $f$ of type $X \Rightarrow Y$, a modulus of continuity, and a proof that $f$ is uniformly continuous with the given modulus. Again, we will leave the projection to the function type implicit and allow us to write $f(x)$ when $f\colon X \to Y$ and $x\colon X$.

## 3.2 Classification of Metric Spaces

There is a hierarchy of classes that metrics can belong to. The strongest class of metrics are the *decidable metrics* where

$$\forall x\,y\,\varepsilon.\, B^X_\varepsilon(x,y) \vee \neg B^X_\varepsilon(x,y).$$

The constructive disjunction here implies there is an algorithm for computing whether two points are within $\varepsilon$ of each other or not. The metric on $\mathbb{Q}$ has this property; however, the metric on $\mathbb{R}$ does not because of the lack of a decidable equality.

The next strongest class of metrics is what I call *located metrics*. These metrics have the property

$$\forall x\,y\,\varepsilon\,\delta.\varepsilon < \delta \Rightarrow B_\delta^X(x,y) \vee \neg B_\varepsilon^X(x,y).$$

This is similar to being decidable, but there is a little extra wiggle room. If $x$ and $y$ are between $\varepsilon$ and $\delta$ far apart, then the algorithm has the option of either return a proof of $B_\delta^X(x, y)$ or $\neg B_\varepsilon^X(x, y)$. This extra flexibility allows $\mathbb{R}$ to be a located metric. Every decidable metric is also a located metric. Some metrics are not located. The standard sup-metric on functions between metric spaces may not be located.

The weakest class of metrics we will discuss are the *stable metrics*. A metric is stable when

$$\forall x\,y\,\varepsilon.\neg\neg B_\varepsilon^X(x,y) \Rightarrow B_\varepsilon^X(x,y).$$

Every located metric is stable. Although we will discuss the possibility of non-stable metrics in section 7, it appears that metric spaces used in practice are stable. This work relies crucially on stability at one point, so we will be assuming that metric spaces are stable throughout this paper.

## 3.3 Complete Metrics

Given a metric space $X$, one can create a new metric space called the *completion of $X$*, or simply $\mathfrak{C}(X)$. The type $\mathfrak{C}(X)$ is defined to be

$$\exists f\colon \mathbb{Q}^+ \Rightarrow X.\forall \varepsilon_1\,\varepsilon_2.B_{\varepsilon_1+\varepsilon_2}^X(f(\varepsilon_1),f(\varepsilon_2))$$

with the ball relation defined to be

$$B_\varepsilon^{\mathfrak{C}(X)}(x,y) := \forall \delta_1\,\delta_2.B_{\delta_1+\varepsilon+\delta_2}^X(x(\delta_1),y(\delta_2)).$$

The definition of $\mathfrak{C}(X)$ may look familiar. It is a generalization of the type that I gave for real numbers in equation 1. In fact, in my actual implementation the real numbers are defined to be $\mathfrak{C}(\mathbb{Q})$.

A complete metric comes equipped with an injection from the original space $unit\colon X \to \mathfrak{C}(X)$ and a function $bind\colon (X \to \mathfrak{C}(Y)) \Rightarrow (\mathfrak{C}(X) \to \mathfrak{C}(Y))$ that lifts uniformly continuous functions with domain $X$ to uniformly continuous function with domain $\mathfrak{C}(X)$. One of the most common way of creating functions that operate on complete metric spaces is by using bind. One first defines a function on $X$, which is easy to work with when $X$ is a discrete space. Then one proves the function is uniformly continuous. After that, bind does the rest of the work.

A second, similar way of creating functions with complete domains is by using $map\colon (X \to Y) \Rightarrow (\mathfrak{C}(X) \to \mathfrak{C}(Y))$. The function map can be defined by $\mathrm{map}(f) := \mathrm{bind}(\mathrm{unit}\circ f)$, but in my implementation, map is more fundamental than bind [7].

I will use the following notation:

$$\begin{aligned} \hat{x} &:= \mathrm{unit}(x) \\ \check{f} &:= \mathrm{bind}(f) \\ \bar{f} &:= \mathrm{map}(f) \end{aligned}$$

The completion operation, $\mathfrak{C}$, and the functions unit and bind together form a standard construction called a *monad* [6]. Monads have been used in functional programs to capture many different computational notions including exceptions, mutable state, and input/output [13]. We will see another example of a monad in section 4.

## 3.4 Product Metrics

Given two metric spaces $X$ and $Y$, their Cartesian product $X \times Y$ forms a metric space with the standard sup-metric:

$$B_\varepsilon^{X\times Y}((x_1,y_1),(x_2,y_2)) := B_\varepsilon^X(x_1,x_2) \wedge B_\varepsilon^Y(y_1,y_2)$$

The product metric interacts nicely with the completion operation. There is an isomorphism between $\mathfrak{C}(X \times Y)$ and $\mathfrak{C}(X) \times \mathfrak{C}(Y)$. One direction I call *couple*. The other direction is defined by lifting the projection functions:

$$\begin{aligned} \text{couple} &: \mathfrak{C}(X) \times \mathfrak{C}(Y) \to \mathfrak{C}(X \times Y) \\ \bar{\pi}_1 &: \mathfrak{C}(X \times Y) \to \mathfrak{C}(X) \\ \bar{\pi}_2 &: \mathfrak{C}(X \times Y) \to \mathfrak{C}(Y) \end{aligned}$$

We denote $\text{couple}(x, y)$ by $\langle x, y \rangle$. The following lemmas prove that these functions form an isomorphism.

$$\begin{aligned} \langle \bar{\pi}_1(z), \bar{\pi}_2(z) \rangle &\asymp z \\ (\bar{\pi}_1 \langle x, y \rangle, \bar{\pi}_2 \langle x, y \rangle) &\asymp (x, y) \end{aligned}$$

## 3.5 Hausdorff Metrics

Given a metric space $X$, we can try to put a metric on predicates (subsets) of $X$. We start by defining the Hausdorff hemimetric. A hemimetric is a metric without the symmetry and identity of indiscernibles requirement. We define the hemimetric relation over $X \Rightarrow \star$ as

$$H_\varepsilon^{X \Rightarrow \star}(\mathcal{A}, \mathcal{B}) := \forall x \in \mathcal{A}. \tilde{\exists} y \in \mathcal{B}. B_\varepsilon(x, y).$$

Notice the use of the classical existential in this definition. In general, we do not need to know which point in $\mathcal{B}$ is close to a given point in $\mathcal{A}$; it is sufficient to know one exists without knowing which one. Furthermore, there are cases when we cannot know which point in $\mathcal{B}$ is close to a given point in $\mathcal{A}$.

This relation is reflexive and satisfies the triangle inequality. It is not symmetric. We define a symmetric relation by

$$B_\varepsilon^{X \Rightarrow \star}(\mathcal{A}, \mathcal{B}) := H_\varepsilon^{X \Rightarrow \star}(\mathcal{A}, \mathcal{B}) \wedge H_\varepsilon^{X \Rightarrow \star}(\mathcal{B}, \mathcal{A}).$$

This relationship is reflexive, symmetric, and satisfies the triangle inequality. Notice that if $\mathcal{B} \subseteq \mathcal{A}$ then $H_\varepsilon(\mathcal{A}, \mathcal{B})$ holds for all $\varepsilon$. The hemimetric captures the subset relationship. If $\mathcal{B} \subseteq \mathcal{A}$ and $\mathcal{A} \subseteq \mathcal{B}$ (i.e. $\mathcal{A} \asymp \mathcal{B}$), then $B_\varepsilon(\mathcal{A}, \mathcal{B})$ holds for all $\varepsilon$. However, axiom 5 for metric spaces requires the reverse implication; if $B_\varepsilon(\mathcal{A}, \mathcal{B})$ holds for all $\varepsilon$, then we want $\mathcal{A} \asymp \mathcal{B}$. Unfortunately, this does not hold in general. Neither does the closedness property required by axiom 4 hold. To make a true metric space, we need to focus on a subclass of predicates that have more structure.

## 4 Finite Enumerations

A finite enumeration of points from $X$ is represented by a list. A point $x$ is in a finite enumeration if there classically exists a point in the list that is equivalent to $x$. We are not required to know which point in the list is equivalent to $x$; we only need to know that there is one. An equivalent definition can be given by well-founded recursion on lists:

$$\begin{aligned} x \in \text{nil} &:= \bot \\ x \in \text{cons}\, y\, l &:= x \asymp y \,\tilde{\vee}\, x \in l \end{aligned}$$

Two finite enumerations are considered equivalent if they have exactly the same members:

$$l_1 \asymp l_2 := \forall x.\, x \in l_1 \Leftrightarrow x \in l_2$$

If $X$ is a metric space, then the space of finite enumerations over $X$, $\mathfrak{F}(X)$, is also a metric space. The Hausdorff metric with the membership predicate defines the ball relation:

$$B_\varepsilon^{\mathfrak{F}(X)}(l_1, l_2) := B_\varepsilon^{X \Rightarrow \star}(\lambda x. x \in l_1, \lambda y. y \in l_2)$$

This ball relation is both closed (axiom 4) and is compatible with our equivalence relation for finite enumerations (axiom 5), so this truly is a metric space.

Finite enumerations also form a monad (I have yet to verify this in Coq). The unit : $X \to \mathfrak{F}(X)$ function creates an enumeration with a single member. The bind : $(X \to \mathfrak{F}(Y)) \Rightarrow (\mathfrak{F}(X) \to \mathfrak{F}(Y))$ function takes an $f: X \to \mathfrak{F}(Y)$ and applies it to every element of an enumeration $l: \mathfrak{F}(X)$ and returns the union of the results.

## 4.1 Mixing Classical and Constructive Reasoning

Proving the ball relation for finite enumerations is closed makes essential use of classical reasoning. Given $\varepsilon$, suppose $B_\delta^{\mathfrak{F}(X)}(l_1, l_2)$ holds whenever $\varepsilon < \delta$. We need to show that $B_\varepsilon^{\mathfrak{F}(X)}(l_1, l_2)$ holds. By the definition of the metric, this requires proving (in part) $\forall x \in l_1. \tilde{\exists} y \in l_2. B_\varepsilon^X(x, y)$. From our assumptions, we know that $\forall x \in l_1. \tilde{\exists} y \in l_2. B_\delta^X(x, y)$ holds for every $\delta$ greater than $\varepsilon$. If we had used a constructive existential in the definition of the Hausdorff hemimetric, we would have a problem. Each different value $\delta$ could produce a *different y* witnessing $B_\delta^X(x, y)$. In order to use the closedness property from $X$ to conclude $B_\varepsilon^X(x, y)$, we need a *single y* such that $B_\delta^X(x, y)$ holds for all $\delta$ greater than $\varepsilon$. Classically we would use the infinite pigeon hole principle to find a single $y$ that occurs infinitely often in the stream of $y$s produced from $\delta \in \{\varepsilon + \frac{1}{n} \,|\, n: \mathbb{N}^+\}$. Such reasoning does not work constructively. Given an infinite stream of elements drawn from a finite enumeration, there is no algorithm that will determine which one occurs infinitely often.

Fortunately, because we used classical quantifiers in the definition of the Hausdorff metric, we can apply the the infinite pigeon hole principle to this problem. We classically know there is some $y$ that occurs infinitely often when $\delta \in \{\varepsilon + \frac{1}{n} \,|\, n: \mathbb{N}^+\}$, even if we do not know which one. For such $y$, $B_\delta^X(x, y)$ holds for $\delta$ arbitrarily close to $\varepsilon$, and therefore $B_\delta^X(x, y)$ must hold for all $\delta$ greater than $\varepsilon$. By the closedness property for $X$, $B_\varepsilon^X(x, y)$ holds as required. The other half of the definition of $B_\varepsilon^{\mathfrak{F}(X)}(l_1, l_2)$ is handled similarly.

Recall that the classical fragment of constructive logic requires that proof by contradiction hold for atomic formulas in order to deduce the rule $\neg\neg \varphi \Rightarrow \varphi$. Because $B_\varepsilon^X(x, y)$ is a parameter, we do not know if it is constructed out of classical connectives. To use the classical reasoning needed to apply the pigeon hole principle, we assume that $\neg\neg B_\varepsilon^X(x, y) \Rightarrow B_\varepsilon^X(x, y)$ holds. This is the crucial point where stability of the metric for $X$ is used.

# 5 Compact Sets

Completing the metric space of finite enumerations yields a metric space of compact sets:

$$\mathfrak{K}(X) := \mathfrak{C}(\mathfrak{F}(X))$$

The idea is that every compact set can be represented as a limit of finite enumerations that approximate it. In order for a compact set to be considered a set, we need to define a membership relation. The membership is not over $X$ because compact sets are supposed to be complete and $X$ may not be a complete space itself. Instead, membership is over $\mathfrak{C}(X)$, and it is defined for $x: \mathfrak{C}(X)$ and $\mathcal{S}: \mathfrak{K}(X)$ as

$$x \in \mathcal{S} := \forall \varepsilon_1\, \varepsilon_2. \tilde{\exists} y \in \mathcal{S}(\varepsilon_2). B_{\varepsilon_1+\varepsilon_2}^X(x(\varepsilon_1), y).$$

A point is considered to be a member of a compact set $\mathcal{S}$ if it is arbitrarily close to being a member of all approximations of $\mathcal{S}$. Thus $\mathfrak{K}(X)$ represents the space of compact subsets of $\mathfrak{C}(X)$.

## 5.1 Correctness of Compact Sets

Bishop and Bridges define a compact set in a metric space $X$ as a set that is complete and totally bounded [1]. In our framework, we say a predicate $\mathcal{A}: X \Rightarrow \star$ is complete if for every $x: \mathfrak{C}(X)$ made from approximations in $\mathcal{A}$, then $x$ is in $\mathcal{A}$:

$$\forall x: \mathfrak{C}(X). (\forall \varepsilon. x(\varepsilon) \in \mathcal{A}) \Rightarrow \exists z \in \mathcal{A}. \hat{z} \asymp x$$

A set $\mathcal{B}\colon X \Rightarrow \star$ is totally bounded if there is an $\varepsilon$-net for every $\varepsilon\colon \mathbb{Q}^+$. An $\varepsilon$-net is a list of points $l$ from $\mathcal{B}$ such that for every $x \in \mathcal{B}$ there (constructively) exists a point $z$ that is constructively in $l$ and $B_\varepsilon(x, z)$. Bishop and Bridges use the strong constructive definition of list membership that tells which member of the list the value is.

$$\forall \varepsilon\colon \mathbb{Q}^+.\, \exists l\colon \text{list } X.\, (\forall x \in l.\, x \in \mathcal{B}) \wedge \forall x \in \mathcal{B}.\, \exists z \in l.\, B_\varepsilon^Y(x, z)$$

Does our definition of compact sets correspond with Bishop and Bridges's definition? The short answer is yes, but there is a small caveat. Our definition of metric space is more general than the one that Bishop and Bridges use. Bishop and Bridges require a distance function $d\colon X \Rightarrow X \Rightarrow \mathbb{R}$. Our more liberal definition of metric space does not have this requirement. I have verified that our definition of compact is the same as Bishop and Bridges's assuming that $X$ is a located metric. If a metric space has a distance function, then it is a located metric. Thus our definition of compact corresponds to Bishop and Bridges's definition of compact for those metric spaces that correspond to Bishop and Bridges's definition of metric space.

It may seem impossible that our definition can be equivalent to Bishop and Bridges's definition when we sometimes use a classical existential quantifier while Bishop and Bridges use constructive quantifiers everywhere. How would one prove Bishop and Bridges version of $x \in \mathcal{S}$ from our version of $x \in \mathcal{S}$? The trick is to use the constructive disjunction from the definition of located metric. Roughly speaking, at some point we need to prove $\exists z \in l.\, B_\varepsilon^Y(x, z)$ from $\tilde{\exists} z \in l.\, B_\varepsilon^Y(x, z)$. This can be done by doing a search though the list $l$ using the located metric property to decide for each element $z_0 \in l$ whether $B_{\varepsilon+\delta}(x, z_0)$ or $\neg B_\varepsilon(x, z_0)$ holds. The classical existence is sufficient to prove that this finite search will successfully find some $z$ such that $B_{\varepsilon+\delta}(x, z)$ holds. The extra $\delta$ can be absorbed by other parts of the proof. The full proof of the isomorphism is too technical to be presented here. A detailed description can be found in my forthcoming PhD thesis or by examining the formal Coq proofs.

## 5.2 Distribution of $\mathfrak{F}$ over $\mathfrak{C}$

The composition of two monads $\mathfrak{A} \circ \mathfrak{B}$ forms a monad when there is a distribution function $\text{dist}\colon \mathfrak{B}(\mathfrak{A}(X)) \to \mathfrak{A}(\mathfrak{B}(X))$ satisfying certain laws [5]. For compact sets, $\mathfrak{K}(X) := (\mathfrak{C} \circ \mathfrak{F})(X)$, the distribution function $\text{dist}\colon \mathfrak{F}(\mathfrak{C}(X)) \to \mathfrak{C}(\mathfrak{F}(X))$ is defined by

$$\text{dist}(l)(\varepsilon) := \text{map}\,(\lambda x.\, x(\varepsilon))\, l.$$

This function interprets a finite enumeration of points from $\mathfrak{C}(X)$ as a compact set. Thus $\mathfrak{K}$ is also a monad (I have yet to verified this in Coq).

## 5.3 Compact Image

We define the compact image of a compact set $\mathcal{S}\colon \mathfrak{K}(X)$ under a uniformly continuous function $\check{f}\colon \mathfrak{C}(X) \to \mathfrak{C}(Y)$ by first noting that applying $f$ to every point in a finite enumeration is a uniformly continuous function, $\text{map}(f)\colon \mathfrak{F}(X) \to \mathfrak{F}(\mathfrak{C}(Y))$. Composing this with dist yields a uniformly continuous function from finite enumeration $\mathfrak{F}(X)$ to compact sets $\mathfrak{K}(Y)$. Using bind, this function can be lifted to operate on $\mathfrak{K}(X)$. The result is the *compact image* function:

$$f \upharpoonright \mathcal{S} := \text{bind}\,(\text{dist} \circ \text{map}(f))(\mathcal{S})$$

Although Bishop and Bridges would agree that the result of this function is compact, they would not say that it is the image of $\mathcal{S}$ because one cannot constructively prove

$$y \in f \upharpoonright \mathcal{S} \Rightarrow \exists x \in \mathcal{S}.\, \check{f}(x) \asymp y.$$

However, I believe one can prove (but I have not verified this yet) the classical statement

$$y \in f \upharpoonright \mathcal{S} \Rightarrow \tilde{\exists} x \in \mathcal{S}.\, \check{f}(x) \asymp y.$$

When $\check{f}$ is injective, as it will be for our graphing example in section 6.1, the constructive existential statement holds.

# 6 Plotting Functions

There are many examples of constructively compact sets. This section illustrates one application of compacts sets, plotting functions.

## 6.1 Graphing Functions

Given a uniformly continuous function $\check{f} : \mathfrak{C}(X) \to \mathfrak{C}(Y)$ and a compact set $\mathcal{D} : \mathfrak{K}(X)$, the graph of the function over $\mathcal{D}$ is the set of points $\{(x, \check{f}(x)) | x \in \mathcal{D}\}$. This graph can be constructed as a compact set $\mathcal{G} : \mathfrak{K}(X \times Y)$. A single point is graphed by the function $g(x) := \langle \hat{x}, f(x) \rangle$. This function is uniformly continuous, $g : X \to \mathfrak{C}(X \times Y)$. The graph $\mathcal{G}$ is defined as the compact image of $\mathcal{D}$ under $g$.

$$\mathcal{G} := g \upharpoonright \mathcal{D}$$

## 6.2 Rasterizing Compact Sets

Given a compact set in the plane $\mathcal{S} : \mathfrak{K}(\mathbb{Q} \times \mathbb{Q})$, we can draw an image of it, or rather we can draw an approximation of it. This process consists of two steps. The first step is to compute an $\varepsilon$-approximation $l := \mathcal{S}(\varepsilon)$. The finite enumeration $l$ is a list of rational coordinates. The next step is to move these points around so that all the points lie on a raster. A raster is simply a two dimensional matrix of Booleans. Given coordinates for the top-left and bottom-right corners, a raster can be interpreted as a finite enumeration. Using advanced notation features in Coq, a raster can be displayed inside the proof assistant. Most importantly, when the constructed raster is interpreted, it is provably close to the original compact set.

## 6.3 Plotting the Exponential Function.

Given a uniformly continuous function $\check{f} : \mathbb{R} \to \mathbb{R}$ and an interval $[a, b]$, the graph of $\check{f}$ over this compact interval is a compact set. The graph is an ideal mathematical curve. This graph can then be plotted yielding a raster that when interpreted as a finite enumeration is provably close to the ideal mathematical curve.

Recall figure 1 from section 1. It is a theorem in Coq that states the (ideal mathematical) graph of the exponential function (which is uniformly continuous on $(-\infty, 1]$) restricted to the range $[0, 3]$ on the interval $[-6, 1]$ is within $\frac{324}{2592}$ (which is equivalent to $\frac{1}{8}$) of the finite set represented by raster shown with the top-left corner mapped to $(-6, 3)$ and the bottom-right corner mapped to $(1, 0)$. The raster is 42 by 18, so, by considering the domain and range of the graph, each pixel represents a $\frac{1}{6}$ by $\frac{1}{6}$ square. The error between the plot and the graph must always be greater than half a pixel. I chose an $\varepsilon$ that produces a graph with an error of $\frac{3}{4}$ of a pixel. In this case $\frac{3}{4} \cdot \frac{1}{6} \asymp \frac{1}{8}$, which is the error given in the theorem.

There is one small objection to this image. Each block in the picture represents an infinitesimal mathematical point lying at the center of the block, but the block appears as a square the size of the pixel. This can be fixed by interpreting each block as a filled square instead of as a single point. This change would simply add an additional $\frac{1}{2}$ pixel to the error term. This has not been done yet in this early implementation.

# 7 Alternative Hausdorff Metric Definitions

There is another possible definition for the Hausdorff metric. One could define the Hausdorff hemimetric as

$$H'_\varepsilon(\mathcal{A}, \mathcal{B}) := \forall x \in \mathcal{A}. \forall \delta. \exists y \in \mathcal{B}. B^X_{\varepsilon + \delta}(x, y).$$

The extra flexibility given by the $\delta$ term also allows one to conclude that there is some $y \in \mathcal{B}$ that is within $\varepsilon$ of $x$ without telling us which one (again, it may be the case that we cannot know which $y$ is the one). Our original definition $H_\varepsilon(\mathcal{A}, \mathcal{B})$ is implied by $H'_\varepsilon(\mathcal{A}, \mathcal{B})$; however, the alternative definition yields more constructive information.

The two definitions are equivalent under mild assumptions. When $X$ is a located metric, then $H'_\varepsilon(\mathcal{A}, \mathcal{B}) \Leftrightarrow H_\varepsilon(\mathcal{A}, \mathcal{B})$. This is very common case and allows us to recover the constructive information in the $H'$ version from the $H$ version.

The constructive existential in the definition of $H'$ would make the resulting metric not provably stable. It is somewhat unclear which version is the right definition for the constructive Hausdorff metric. The key deciding factor for me was that I had declared the ball relation to be in the `Prop` universe. Coq has a `Prop`/`Set` distinction where values in the `Prop` universe are removed during program extraction [11]. To make program extraction sound, values outside the `Prop` universe cannot depend on information inside the `Prop` universe. This means that even if I used the $H'$ definition in the Hausdorff metric, its information would not be allowed by Coq to construct values in `Set`. For this reason, I chose the $H$ version with the classical quantifiers for the definition of the Hausdorff metric. Values with classical existential quantifier type have no information in them and naturally fit into the `Prop` universe.

## 8  Conclusion

This work shows that one can compute with and display constructively compact sets inside a proof assistant. We showed how to graph uniformly continuous functions and render the results. We have turned a proof assistant into a graphing calculator. Moreover, our plots come with proofs of (approximate) correctness.

Even though a classical quantifier in the Hausdorff metric is used, it does not interfere with the computation of raster images. This development shows that one can combine classical reasoning with constructive reasoning. The classical existential quantifier was key in allowing us to use the pigeon hole principle to prove the closedness property of the Hausdorff metric.

All of the theorems in this paper have been verified by Coq except where indicated otherwise. Those few theorems that have not been verified in Coq are not essential and have not been assumed in the rest of the work (the statements simply do not appear in the Coq formalization). This formalization will be part of the next version of the CoRN library [3], which will be released when Coq 8.2 is released.

Given my previous work about metric spaces and uniformly continuous functions [8], the work of defining compact sets and plotting functions took only one and a half months of additional work.

This work provides a foundation for future work. One can construct more compact sets such as fractals and geometric shapes. Proof assistants could be modified so that the high resolution display of a monitor could be used instead of the "ASCII art" notation that is used in this work.

## 9  Acknowledgments

I would like to thank my advisor, Bas Spitters, for suggesting the idea that compact sets can be defined as the completion of finite sets.

I would also like to thank Jasper Stein whose implementation of Sokoban in Coq [10] using ASCII art inspired the idea of using Coq's notation mechanism for displaying graphs inside Coq.